# A Thin Film Broadband Absorber Based on Multi-sized Nanoantennas

Yanxia Cui[1, 2], Jun Xu[1], Kin Hung Fung[1], Yi Jin[2,*], Anil Kumar[1],
Sailing He[2], and Nicholas X. Fang[1,*]

[1] Department of Mechanical Science and Engineering and Beckman Institute of Advanced Science and Technology, University of Illinois at Urbana-Champaign, Urbana, Illinois 61801, USA
[2] Centre for Optical and Electromagnetic Research, State Key Laboratory of Modern Optical Instrumentation, Zhejiang University; Joint Research Centre of Photonics of the Royal Institute of Technology (Sweden) and Zhejiang University, Zijingang Campus, Zhejiang University, Hangzhou 310058, China
*Corresponding author: yinjin@coer.zju.edu.cn, nicfang@mit.edu

We experimentally demonstrate an infrared broadband absorber for TM polarized light based on an array of nanostrip antennas of several different sizes. The broadband property is due to the collective effect of magnetic responses excited by these nano-antennas at distinct wavelengths. By manipulating the differences of the nanostrip widths, the measured spectra clearly validate our design for the purpose of broadening the absorption band. The present broadband absorber works very well in a wide angular range.

In the last decade, plasmonic nano-antennas have experienced a drastical boom period due to their enormous capability to compress light into a subwavelength region with an extremely strong amplitude.[1,2] To date, they have found significant application in diverse areas including sensor detection,[3] solar power harvesting,[4] thermal emission,[5] biomedical imaging,[6] ultrafast modulating,[7] etc. Patterned plasmonic antennas play a significant role for the design of thin film light absorbers, which suppress both the transmission and the reflection while maximizing the absorption. The first perfect absorber that composed by metallic split ring resonators and cutting wires was demonstrated by Landy *et al.*[8] Then, it was followed by some work to improve the angular and polarization performance.[9-12]

Nevertheless, all of the above absorbers work at a single band frequency which limits the pragmatic applications such as THz multi-frequency spectroscopy detection.[13] By incorporating different patterns of metallic elements, two dual band absorbers were carried out by different groups.[14,15] Recently, it was reported that based on an H-shaped nano-resonator array, a dual band plasmonic metamaterial absorber could also be constructed.[16] But they are still limited to a relative narrow band response. So far, to design a thin film absorber with broadband spectrum is still quite challenge. In our group, we have made some efforts in this aspect, by stacking multiple layers of metallic crosses with different geometrical dimensions to merge several closely positioned resonant peaks in the absorption spectrum.[17] However, this proposal suffers from one crucial drawback, namely that in the fabrication it is difficult to obtain perfect alignment to match the relative position of each pattern in different layers.

It is well known that a three layered structure composed by an array of plasmonic nanostrip antennas of a fixed width on top of a ground reflecting mirror and a very thin spacer layer[18] can efficiently absorb electromagnetic wave at a certain frequency. The principle of the light absorbing is that the upper strip and the ground metal layer support a pair of anti-parallel dipoles with quite closed distance in-between, the interference of those two dipoles in far field is destructive due to their π shift phase difference so that the reflection can be totally cancelled.

In this letter, also aiming at broadening the absorption band, we borrow the concept of the collective effect of multiple different oscillators[19] and design a broadband light absorber for TM polarized light at infrared regime by tiling an array of multi-sized plasmonic strip antennas on top of a back metallic mirror; see its three-dimensional (3D) and two-dimensional (2D) (*x-z* plane) configurations in Fig. 1(a) and its inset, respectively. It is noted that for such a broadband absorber, the total thickness of the device is still maintained in the deep subwavelength scale that is same as the traditional single band absorber.

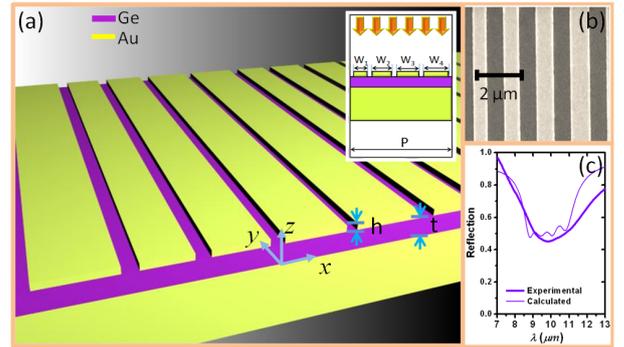

Fig. 1. (Color online) Schematic diagram of the proposed three-layered light harvesting absorber. $P = 6.06$ *μm*, $t = 340$ *nm*, and $h = 16.5$ *nm*. The widths of the top strips (from $W_1$ to $W_4$) form an arithmetic series, with $W_3 = 830$ *nm* and the difference $\Delta = 85$ *nm*. (b) The SEM image of our fabricated sample. (c) The experimental (thick) and simulation (thin) results of the reflection spectrum for the sample presented in (b).

Such a three-layered sample is easy to fabricate. 100nm thick gold film was deposited on the clean glass substrate by E-beam evaporation, and 340nm germanium film was followed by using same method. Within the considered wavelength range from 7 to 13 *μm*, the refractive index of germanium is $n = 4$. Employing E-beam lithography and lift-off technique, the gold nanostrip antennas of four different widths (labelled by $W_{1-4}$) within a regular period $P = 6.06$ *μm* were patterned on the top. The height of strips is very small ($h = 16.5$ *nm*), that was measured by the atomic force microscopy (AFM) method. $W_{1-4}$ forms an arithmetic progression with a fixed $W_3$ (equalling to 830 *nm*) and a tuning difference ($\Delta$). The spaces between neighbouring strips were kept identical. The geometrical parameters of the fabricated samples were determined by scanning electron microscope (SEM) imaging. Fig. 1(b) shows a typical region of a sample surface with $\Delta = 85$ *nm*. In practice, dimensions of the fabricated samples suffer $\pm 10$ *nm* deviation from the target designs on *x-y* plane due to the limitation of fabrication techniques.

Since the ground gold film blocks all light transmission ($T = 0$), the absorption $A$ can be calculated based on reflection ($R$) by $1 - R$. We used the Fourier Transform Infrared Spectroscopy (FTIR) method to characterize the overall reflection spectrum at normal incidence. It is noted that our proposal is a 2D structure which has the resonant response only for TM polarized light (the magnetic field is perpendicular to the x-z plane, defined as $H_y$); hence, only 50% energy of the non-polarized light can response with our device. For the sample in Fig. 1(b), we plot its measured reflection spectrum in Fig.1 (c) (thick), which is consistent with its simulated result (thin) in terms of the reflectivity and the bandwidth. The credibility of measurement can be further confirmed based on the study of strip width differences ($\Delta$); shown in Fig. 4(b). Our simulation is based on the Rigorous Coupled-wave Analysis (RCWA) method[20] and its validity of our simulation has already been proved.[21] The permittivity of gold at wavelengths shorter than 9 μm are from Ref. 22, while others out of that range are obtained by linear interpolation. In addition, other noble metals are also applicable for the current design due to their very similar electronic properties in this frequency range.

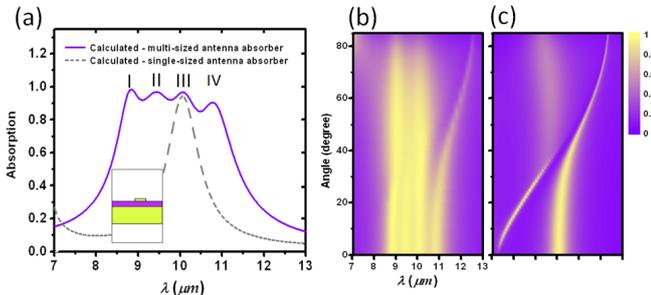

(Color online) (a) Comparison of the absorption spectra between the multi-sized antenna and the single-sized antenna absorbers; (b) and (c) are their angular spectra, respectively.

The superiority of our idea can be clearly shown by comparing it with the absorber of a single-sized nanostrip antenna array. For example, we studied the control case that reserved only the nanostrip of width $W_3$, as shown by the inset of Fig. 2(a). In Fig. 2(a), we plot the calculated absorption spectra for the control case (dashed) under TM polarized incidence. Evidently, in the case of our multi-sized antenna absorber, the absorbing spectrum is very broad as a result of the merging of four close peaks (I, II, III and IV), of which peak III is located closely to the resonant peak of the single antenna absorber. By calculation, we know that our absorber has a FWHM equal to 31.3% at the centred wavelength $\lambda = 9.84$ μm; for the single-sized antenna absorber, FWHM is only 9.2%, which is less than one third of our proposal. Another evaluation criteria is the total absorbed energy efficiency ($\eta$); we check the photon energy range [0.095, 0.18] eV, i.e., the wavelength range [7, 13] μm. For our broadband absorber, $\eta$ is 55.6%, which is about twice as high as that of the control case, where $\eta$ equals 23.1%.

In Fig. 3(a), we plot the normalized magnetic field ($H_y$) (left) and current density (right) distributions at the resonant peak $\lambda = 10.05$ μm for the single-sized antenna absorber. From this figure we clearly see that light is strongly concentrated in the dielectric core region underneath the gold nanostrip and antiparallel currents are excited in the upper nanoantenna layer and the ground metal film. This is indeed a magnetic resonance, similar to that in Ref. 3. We note that the geometrical parameters, like the dielectric thickness and the periodicity, are important factors for designing efficient absorbers. It is because such parameters can influence the impedance of the whole structure $z = \mu_{eff}/\varepsilon_{eff}$, where $\mu_{eff}$ and $\varepsilon_{eff}$ are the effective permittivity and permeability of the bulk multilayer structure, while perfect absorption only happens when $z$ matches that of vacuum[23]. We can also consider the ground gold film as a mirror to generate an image of the upper nanostrip object. We composed a structure by removing the ground gold film and adding corresponding dielectric and nanostrip layers as shown by the inset in Fig. 3(b). By simulating its field distribution at $\lambda = 10.05$ μm, we see that the $|H_y|$ distribution is in symmetry with respect to $z = 0$ plane and it has exactly the profile as that in Fig. 3(a) in $z > 0$ region. By observing the current distribution in Fig 3(b), antiparallel currents in two strip-antennas can be clearly identified, which is responsible for the strong magnetic resonance in the core region.

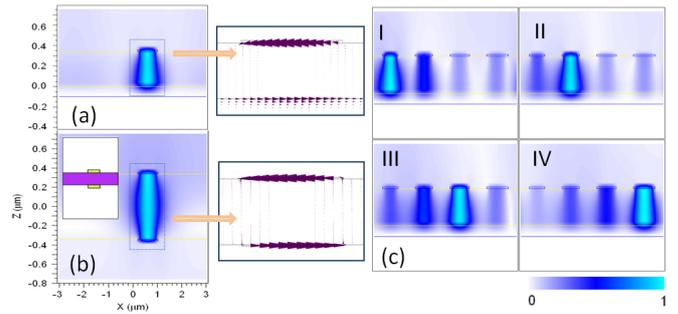

Fig. 3. (Color online) (a) Normalized magnetic field (left) and current density distributions (right) of the single-sized antenna absorber at $\lambda = 10.05$ μm; (b) Field distributions for a related structure with the inset showing its configuration at $\lambda = 10.05$ μm; (c) Field distributions of our multi-sized antenna absorber at its peaks.

At this point, it is necessary to refer readers to a work done by Hibbins *et. al*[18] in 2004, which concerns the microwave absorption of a scale-up sample. In this literature, fraction ratio ($F$), defined by the strip width over the period ($W/P$), is very large ($F = 95\%$); hence the gap between neighbouring strips are quite tiny. It is mentioned that the explanation of the physical mechanism for such a resonating structure is misleading by saying that half of the standing wave was compressed in the tiny gap region.[18] We studied the situation after widening the air gap between the strips, and found that the same resonance with a very good absorption can still be excited with results similar to what we have achieved in Fig. 3(a). However, the separation between neighbouring strips is limited because if the period approaches the considered wavelength range, surface plasmon polaritons (SPP) resonance[1] arises and will deteriorate the localized resonating mode under strips. Therefore, the absorption efficiency at resonance will drop sharply.

We also want to emphasize that the nanostrip absorber with a small fraction ratio owns a large absorption cross section, so that we can take advantage to make efficient broadband absorbers. In the other words, the relatively large spare region between neighbouring strips offers us possibilities to disperse more antennas resonating at other wavelengths in the upper layer, which could be one solution to broaden the absorption band; this is why our idea in Fig. 1(a) comes into being. Based on such an idea, we manage to obtain the expected results shown in Fig. 2(a). In Fig. 3(c) we plot the field distributions at the four peaks of the thick curve in Fig. 2(a), respectively; we find that the energy centre shifts from the region under the first strip to the fourth when we tune the wavelength from peak I to peak IV. We note that based on such means, the broadening effect is restricted due to the existence of SPP band.

In Fig. 2(b) and (c), we show the simulated angular spectra of the multi-sized and single-sized antenna absorbers, respectively.

Due to the localized plasmonic resonance with strong magnetic field excited in a subwavelength region, the absorber should be angular insensitive, similar as that in Ref. 3. However, for the present single-sized antenna structure, since its periodicity is close to the wavelength instead of subwavelength, as we mentioned before, SPP resonance[1] arises. This SPP resonance brings significant interaction with the localized mode of the nanostrip antenna; hence we see the mode splitting happens at the incident angle $\theta = 40º$ in Fig. 2(c). However, for our multi-sized antenna absorber, the interaction between the SPP mode and the localized mode becomes much weaker. In Fig. 2(b), it is seen that the spectrum band width keeps well with varied incident angles; e.g., at $\theta = 50º$, our absorber still has FWHM equal to 28.2% at the centred wavelength $\lambda = 9.55\ \mu m$. This is because the additional nanostrips of non-uniformed sizes can undermine the periodical distributing effect of strips. We note that since the structure still holds a periodicity of $P = 6.06\ \mu m$, the energy coupling to the SPP band for peak IV is still notable in Fig. 2(b).

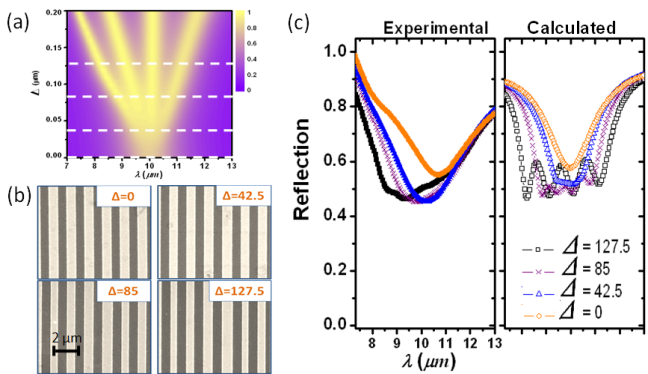

Fig. 4. (Color online) (a) Absorption spectra when tuning the difference ($\varDelta$) from 0 to 200 nm with four dashed lines denoting the cases when $\varDelta = 0$, 42.5, 85 and 127.5 nm. For the four cases, we show the SEM images (b), experimental (thick) and simulated (thin) overall reflection spectra (c).

Last, we investigated the influence of the strip width difference ($\varDelta$) on the absorbing performance while keeping other parameters the same as those in Fig. 1(a). The absorption spectra as $\varDelta$ varies from 0 to 200 *nm* are shown in Fig. 4(a) with the dashed lines denoting four cases when $\varDelta = 0$, 42.5, 85 and 127.5 *nm*. We see that when the four strips have the same width, i.e., $\varDelta = 0$, the absorption spectrum is the narrowest and the absorption peak is the lowest; hence the overall absorbing efficiency is the smallest. As we augment the difference, the absorption spectrum becomes wider with higher absorbtivity; further increasing the difference will make the structure acting as an efficient multi-band absorber. It is easy to explain these features. For the case of $\varDelta = 0$, the band is narrow because all the strips resonate at the same wavelength, and the efficiency is poor because the impedance is mismatched due to the periodicity of the strips has been changed to a quarter of its original value[24]. If we tune the spacer thickness $t = 220\ nm$, full absorption can be obtained at its peak wavelength. When we introduced a large enough width difference into the four strips in each period, e.g., for $\varDelta > 50$ nm, those strips will perform as distinct resonators working at four different wavelengths so that they go back to their original states of periodicity $P = 6.06\ \mu m$ to refit the impedance match condition as the single strip antenna case which was already shown in Fig. 2(a) by the dashed line. When the strip width difference is not too large to make the four corresponding resonating wavelengths totally separated, a continuous broadband efficient absorbing spectrum can be formed just as what we have shown in Fig. 2(a) by solid line and also in Fig. 4(a) when $\varDelta = 85$ nm. We fabricated samples at these four cases with their SEM images shown in Fig. 4(b) and measured the overall spectra using FTIR, which are shown in Fig. 4(c) left with a comparison of the simulated results in Fig. 4(c) right. It is evidently seen that we have successfully reproduced the above-mentioned features, i.e., both the spectrum band and the absorption strength. The spectrum distortion for the cases $\varDelta = 85$ and 127.5 nm is reasonable because of the dimension roughness generated during the fabrication process.

In conclusion, we have reported our recent experimental progress of designing a broadband absorber based on multi-sized nanostrip antennas at the infrared regime for TM polarization. In the future, further work can be done to evaluate the randomness and the packing numbers of the strip antennas.

We appreciate Rohith Reddy from Prof. Rohit Bhargava's group at UIUC for characterizing the spectrum by FTIR and Hyungjin Ma for measuring the strip height by AFM. Y. Cui would thank Jianwei Tang and Yingran He for helpful discussions. This work is partially supported by the National Science Foundation (CMMI 0846771), AOARD, and the National Natural Science Foundation of China (60990320 and 60901039).